\documentclass[useAMS,usenatbib,usegraphicx]{mn2e}

\usepackage[letterpaper, totalwidth=480pt, totalheight=680pt]{geometry}

\begin{document}

\title { Astrometric effects of non-uniform telescope throughput } 

\author[M. Gai and R. Cancelliere]{M. Gai$^{1}$ 
\thanks { E-mail: gai@oato.inaf.it (MG), cancelli@di.unito.it (RC)} 
and R. Cancelliere$^{2}$ \\ 
$^{1}$Istituto Nazionale di Astrofisica - Osservatorio Astronomico di 
Torino, V. Osservatorio 20, 10025 Pino T.se (TO), Italy \\ 
$^{2}$Dipartimento di Informatica, Universit\`a di Torino, 
C.so Svizzera 185, 10149 Torino, Italy }

\date{ Accepted 2008 September 22.  Received 2008 September 22; in original 
form 2008 August 22 }

\pagerange{\pageref{firstpage}--\pageref{lastpage}} \pubyear{2007}

\maketitle

\label{firstpage}

\begin{abstract}
In real telescopes, the optical parameters evolve with time, and 
the degradation is often not uniform. 
This introduces variations in the image profile and therefore 
photo-centre displacements which, unless corrected, may result in 
astrometric errors. 
The effects induced on individual telescopes and interferometric 
arrays are derived by numerical implementation of a range of cases. 
The results are evaluated with respect to the potential impact on the 
most relevant experiments for high precision astrometry in the near 
future, i.e. Gaia, PRIMA and SIM, and to mitigation techniques applicable 
from design stage to calibrations. 
\end{abstract} 


\section{Introduction}
\label{intro}

The mirrors of ground based instruments suffer degradation of the 
reflecting surface, e.g. by oxidation or contamination of the coating, 
due to dust, moisture and chemical agents in the environments 
\citep{Vucina}. 
The result is a degradation of the optical throughput, which in general is 
not uniform, but rather described by a complex, often patchy, distribution. 
Also, the overall telescope transmission is progressively reduced, 
so that it becomes necessary to process in particular the primary 
mirror regularly for removal of the degraded reflecting coating and 
application of a new one. 

Space optics is expected to be much less affected by surface characteristics 
variations, thanks to the stable environment and suitable protective layers, 
but degradation e.g. due to $\gamma$ radiation is still experienced, up to 
1\% of the overall throughput, depending on the choice of substrate and 
coating material \citep{Baccaro}. 

Even small variations are still potentially relevant with respect to 
high precision astrometric measurements, and as such their effects 
have been investigated and are discussed in this document. 
The main experiments in which astrometric errors induced by 
non-uniform optical throughput variation are potentially relevant 
are Gaia, PRIMA and SIM. 

Gaia \citep{Gaia} is the most important European space mission devoted 
to micro-arcsec ($\mu$as) astrometry currently being implemented. 
Interferometry also aims at comparable levels of precision by means of 
both ground based arrays, e.g. the Phase Referenced Imaging and 
Microarcsecond Astrometry [PRIMA] facility \citep{PRIMA} of the Very 
Large Telescope Interferometer [VLTI] of the European Southern 
Observatory [ESO], or from space, e.g. the Space Interferometer Mission 
[SIM] \citep{SIM}. 

The astrometric performance of an astronomical instrument depends on 
the detected signal profile. 
The Point Spread Function (PSF), i.e. the intensity distribution from 
a point-like object at infinity, is built from the diffraction integral, 
described in optics textbooks, e.g. \citet{born}; the numerical 
implementation used here was described in a previous paper 
\citep{gai07}. 
The PSF includes, by a suitable wavefront error (WFE) map, 
the realistic description of several optical characteristics. 
The polychromatic PSF is produced by superposition, 
weighted by spectral distribution, of the monochromatic PSFs. 

Hereafter, $I_m$ is the monochromatic PSF at a given wavelength $\lambda$, 
and $E_m$ is the corresponding complex amplitude distribution on the focal 
plane; $F$ is the effective focal length of the telescope. 
The coordinates on pupil and focal plane are respectively $\{\xi,\,\eta\}$ 
and $\{x,\,y\}$ (linear units). 
The complex amplitude depends on the pupil function $P$ and on the aperture 
transmission function $T$: 
\begin{equation}
\label{eq:ampli_mono}
E_m \left( x, y \right) = 
k \int d\xi \, d\eta \, 
T \left( \xi, \eta \right) \cdot P \left( \xi, \eta \right) \ 
e^{-i\pi \left(x \xi + y \eta\right) / \lambda F} \, . 
\end{equation}
The constant $k$ provides the appropriate photometric result associated 
with the source emission, exposure time and collecting area. 
The monochromatic PSF can be expressed as 
\begin{equation}
\label{eq:PSF_mono}
I_m \left( x, y \right) = \left| E_m \left( x, y \right) \right| ^{2} \, . 
\end{equation}

The pupil function 
$P \left( \xi, \eta \right) = e^{i \Phi \left( \xi, \eta \right)}$  
depends on the phase aberration function $\Phi$, often expanded 
e.g. in terms of the Zernike functions $\Phi_n$ \citep{born}: 
\begin{equation}
\label{eq:aberr}                                                                                       
\Phi \left( \rho ,\theta \right) = \frac{2\pi}{\lambda} WFE = 
\frac{2\pi}{\lambda} \sum_n  A_n \varphi_n 
\left( \rho, \theta \right) \ . 
\end{equation} 
The aperture transmission function $T$ is often considered to take unity 
value inside the pupil and zero outside; it is therefore used simply as 
a convenient numerical tool for insertion of the pupil geometric 
description. 
In the current simulation, however, it is used also to describe the variation 
of optical throughput of the telescope throughout the pupil. 
The formalism is similar to that adopted by \citet{Linfield}, in the framework 
of straylight analysis for a Terrestrial Planet Finder (TPF) coronagraph. 
The most relevant changes are a simplified treatment, since we consider 
only transmission variations and neglect phase contributions, and the focus 
on astrometric effects, i.e. image photo-centre displacements. 

We expect that a uniform variation of the optical throughput over the 
whole telescope aperture does not introduce any variation on the image 
photo-centre; however, for a generic aberrated configuration, a 
non-uniform pupil transmission introduces an amplitude modulation of 
the pupil function, thus modifying the resulting focal plane energy 
distribution, hence the measured signal profile, and consequently 
its photo-centre estimated position. 
Such photo-centre displacement is associated with an error on 
the astrometric measurement. 
Although in general the optical throughput is wavelength dependent, 
our analysis is referred to a single spectral type source, and uniform 
spectral dependence is assumed. 
Further levels of detail may be introduced in future investigations. 

In section \ref{sec_1tel} we discuss the introduction of transmission 
non-uniformity features and evaluate a set of reference cases focused 
on the astrometric performance of one telescope. 
In section \ref{sec_interf} we analyse some consequences on the measurements 
of an interferometric array. 
Then, in section \ref{sec_discuss}, we discuss the implications of our 
findings on design and calibration of astrometric instruments. 
Finally, we draw our conclusions. 

\section{Single telescope}
\label{sec_1tel}

The analysis is focused on a configuration similar to the Gaia telescope 
optics and operation, also because its rectangular geometry is marginally 
more convenient for numerical simulation.  
The primary mirror size is $1.45 \times 0.45\,$m; the effective focal length 
is $EFL = 35\,$m, resulting in an optical scale of 5.89~arcsec/mm. 
%
\begin{figure}
\includegraphics[width=84mm]{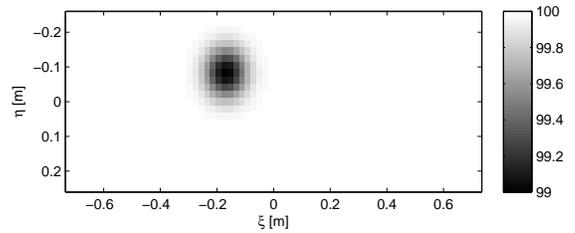}
\caption { Throughput distribution (in percent) for a selected patch 
position. The transmission is reduced locally by $\le$1\%. } 
\label{fig_patch}
\end{figure}
For simplicity, the study case in this document is one-dimensional; 
this is also  most immediately related to the Gaia elementary measurements. 
The PSF is integrated in the direction not used for the astrometric 
measurement by binning on the CCD detector; this is simulated by numerical 
integration over an appropriate region. Also, some of the effects associated 
with the instrument characteristics and operations are included in the form 
of equivalent effects on the PSF through the corresponding Modulation Transfer 
Function (MTF); in particular, this includes the detector pixel 
(size $10\times30\,\mu$m) and Time Delay Integration (TDI) readout 
\citep{pasp98}. 
The values used are representative of the Gaia nominal configuration and 
operation. 

The optical throughput perturbations described below are applied to a range 
of cases with different image quality, achieved by selection of 
the 21 lowest order Zernike coefficients, corresponding to the normalised 
radius raised up to the fifth power, as in the table included in 
\citet{gai05}. 
The optical images are generated as pseudo-polychromatic, using a 
central wavelength of 617~nm, with weight 100\%, and two side wavelengths 
at $\pm$100~nm, with weight 50\%, approximating the instrument response 
to a blackbody source at $T = 3 \times 10^{4}\,$K. 
The photo-centre estimation is performed using the centre of gravity 
algorithm, as it is simple and independent of signal models. 

\begin{figure}
\includegraphics[width=84mm]{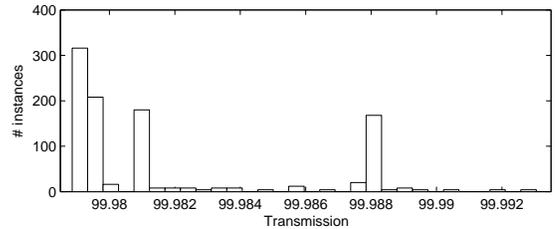}
\caption { Histogram of throughput (in percent) vs. patch position ranging 
over the telescope pupil. } 
\label{fig_transm_hist_1}
\end{figure}
\subsection{Opacity patches}
The telescope throughput degradation is introduced as a localised reduction 
with respect to the nominal unity value ($T_0 = 1$), i.e. an opacity patch 
$p$, so that the local transmission over the pupil is $t \left( \xi, \eta 
\right) = T_0 - p \left( \xi, \eta \right)$. 
A Gaussian shape of the opacity patch is adopted, with given characteristic 
size $\sigma$ ($\sigma = 5\,$cm for most of the cases below), peak position 
$\{\xi_0, \eta_0\}$ and peak value $p_0$: 
\begin{equation}
\label{eq:patch}
p \left( \xi, \eta \right) = p_0 \cdot \exp 
\left[ -{ (\xi - \xi_0)^2 + (\eta - \eta_0)^2 \over 2 \sigma^2 }
\right] \, . 
\end{equation}
For all cases investigated, the peak value is set to $p_0 = 1\%$, i.e. 
the perturbed telescope transmission ranges between 99\% and 100\% of 
the unperturbed value. 
The opacity distribution is therefore not normalised, and its area is 
$A_p = 2 \pi \sigma^2$. A selected case corresponding to patch position 
close to the centre of the pupil is shown in figure \ref{fig_patch}. 
The overall optical throughput variation can be expected to be dependent 
on the peak and size of the opacity distribution, i.e. of order of 
$p_0 \cdot A_p / A_T$, where $A_T = L_\xi L_\eta$ is the primary mirror 
area. 
For a generic centre position, part of the patch is outside the pupil, 
further reducing the throughput degradation. 
The overall transmission variation, averaged over the pupil, is 
thus well below 0.1\%, because the opacity patch is much smaller than 
the primary mirror size. 
Thus, photometric measurements are affected by marginal degradation. 

\subsection{Simulation}
The position of the opacity patch is translated uniformly over 100 position 
on the long side of the telescope, and over 10 positions on the short side; 
each case is thus evaluated on a total number of 1000 instances. 
The pupil and focal plane sampling resolution is respectively 
$2 \times 2\,$cm and $2 \times 6\,\mu$m. 
An histogram of the transmission distribution is shown in figure 
\ref{fig_transm_hist_1}. 
The throughput mean value over the sample of patch positions is 
0.99982, with standard deviation $3 \times 10^{-5}$. 
Hereafter, a range of WFE cases is evaluated against the set of throughput 
perturbations described above. 
In all cases, the readout area is set to $12 \times 12$ pixels. 

\subsubsection{Case 1} 
\begin{figure}
\includegraphics[width=84mm]{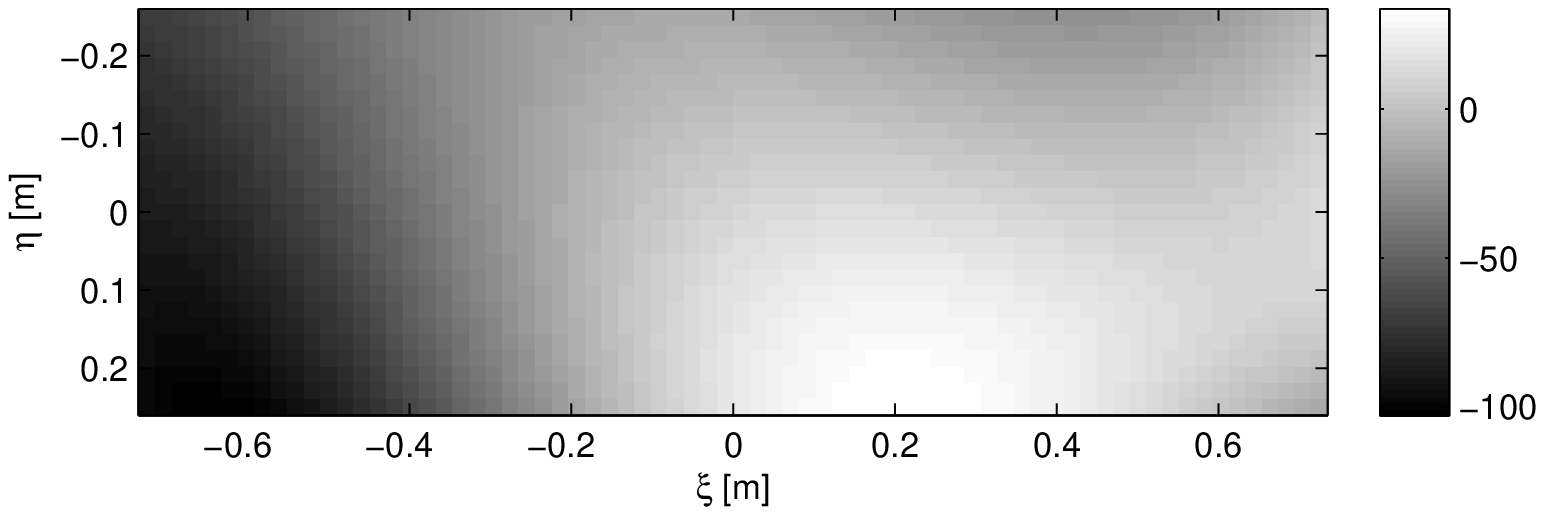}
\includegraphics[width=84mm]{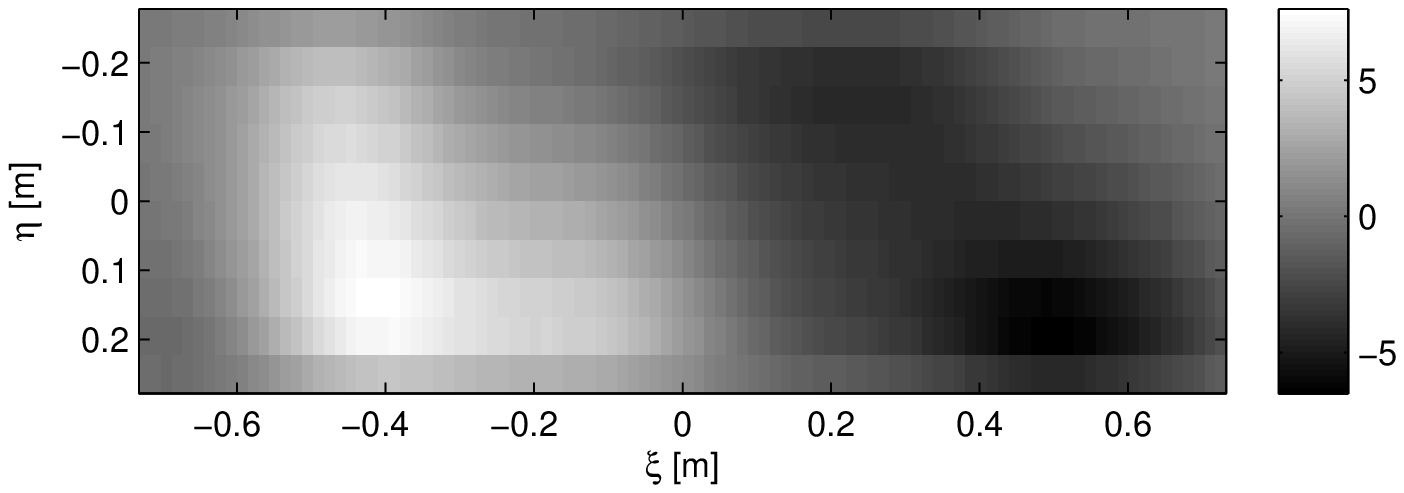}
\includegraphics[width=84mm]{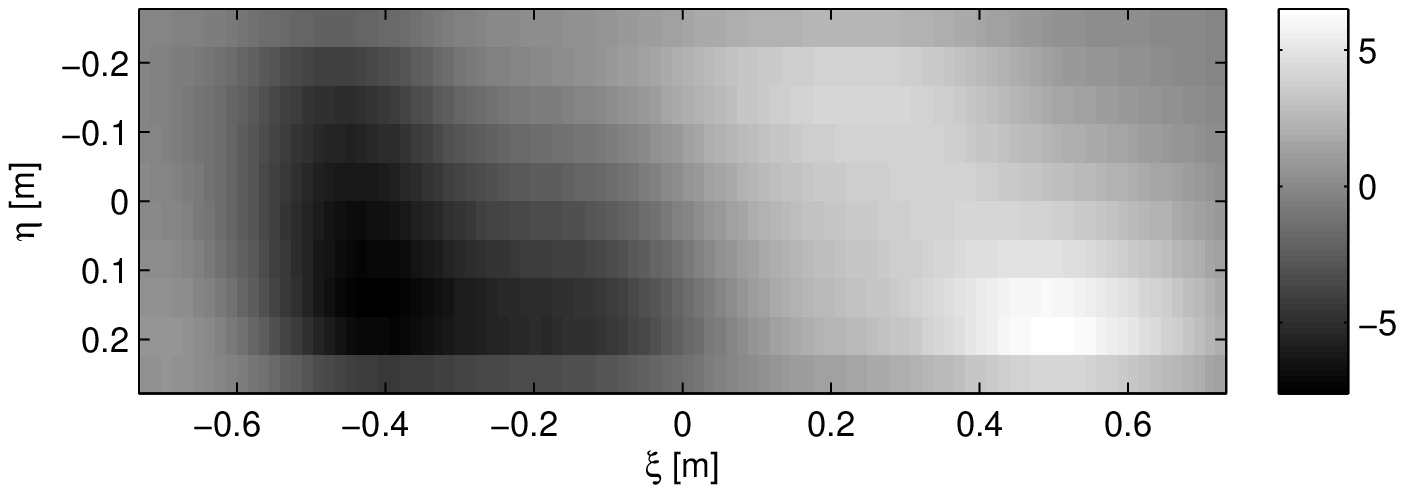}
\caption { Map of WFE for case 1 (top, in nm); distribution of photo-centre 
(in $\mu$as) vs. opacity patch position, case 1 (mid) and 4 (bottom, 
aberration reversed in sign). } 
\label{fig_wfe_cog_c1_c4}
\end{figure}

A set of random values of the 21 lowest order Zernike coefficients is selected, 
with normal distribution and peak value of 50~nm. 
The corresponding WFE has RMS value of 35.05~nm, corresponding to a fairly 
good optical quality, $\sim\lambda/20$ at $\lambda = 700\,$nm. 
The WFE distribution over the pupil is shown in figure \ref{fig_wfe_cog_c1_c4} 
(top). 
The photo-centre computed on each instance of opacity patch position 
is affected by displacements with respect to the unperturbed case; its 
distribution vs. opacity patch position is shown in figure 
\ref{fig_wfe_cog_c1_c4} (mid). 
The photo-centre displacement mean and RMS are respectively 0.03 
and 3.34~$\mu$as; the peak-to-valley (PTV) is 14.13~$\mu$as. 

\subsubsection{Case 2} 
In order to evaluate the dependence on the amount of aberration, the 
WFE shape is retained, but its amplitude is reduced by a factor two by 
halving all Zernike coefficients. 
The resulting RMS WFE is thus reduced to 17.53~nm; the map is not shown 
here, since with respect to figure \ref{fig_wfe_cog_c1_c4} (top) only the 
scale changes, by a factor two. 
The photo-centre distribution is similar to that from case 1 
(figure \ref{fig_wfe_cog_c1_c4}, mid), but it is also reduced by 
a factor two; the mean and RMS are respectively 0.02 and 1.67~$\mu$as; 
the PTV is 7.09~$\mu$as. 
The photo-centre discrepancy between case 1 and case 2, after multiplication 
of the latter by a factor two, have mean and RMS respectively zero and 
0.03~$\mu$as. 

\subsubsection{Case 3} 
Similarly, in case of larger aberration with the same shape (achieved by 
multiplication by a factor two of the Zernike coefficients, resulting in 
RMW WFE = 70.11~nm), the photo-centre displacement is doubled, but again 
similar to that of case 1 (figure \ref{fig_wfe_cog_c1_c4}, mid). 
The mean and RMS photo-centre displacement are respectively 0.07 and 
6.67~$\mu$as; the PTV is 27.92~$\mu$as. 
The photo-centre discrepancy between case 1 and case 3, after division 
of the latter by a factor two, have mean and RMS respectively zero and 
0.09~$\mu$as. 
\begin{figure}
\includegraphics[width=84mm]{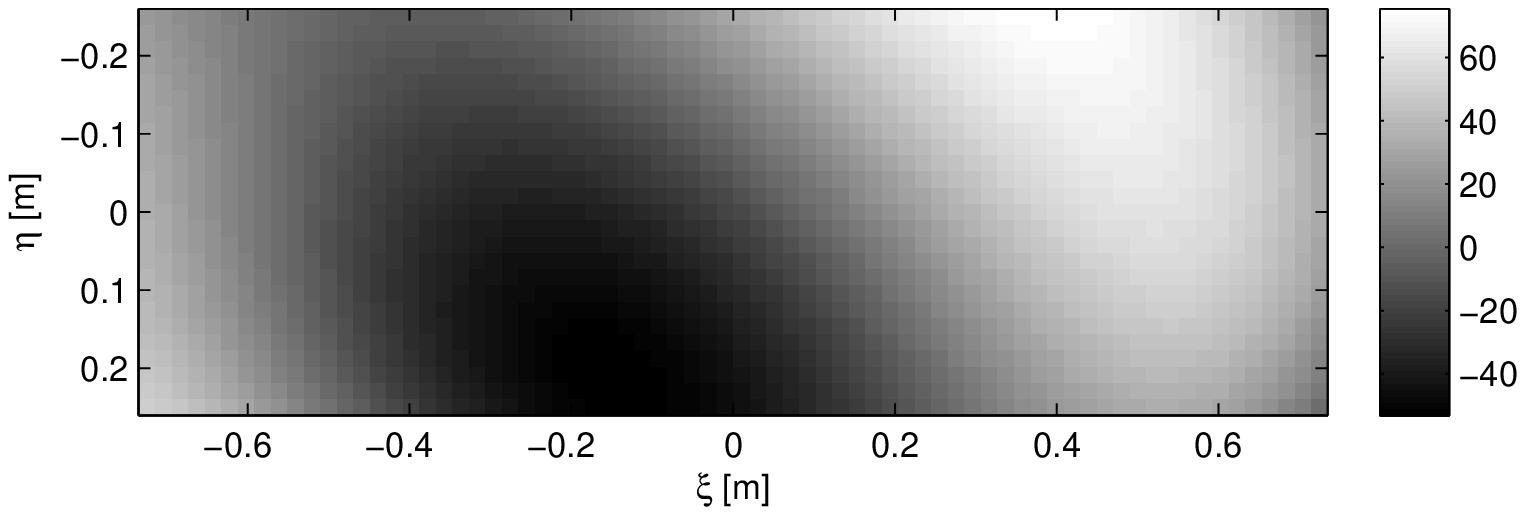}
\includegraphics[width=84mm]{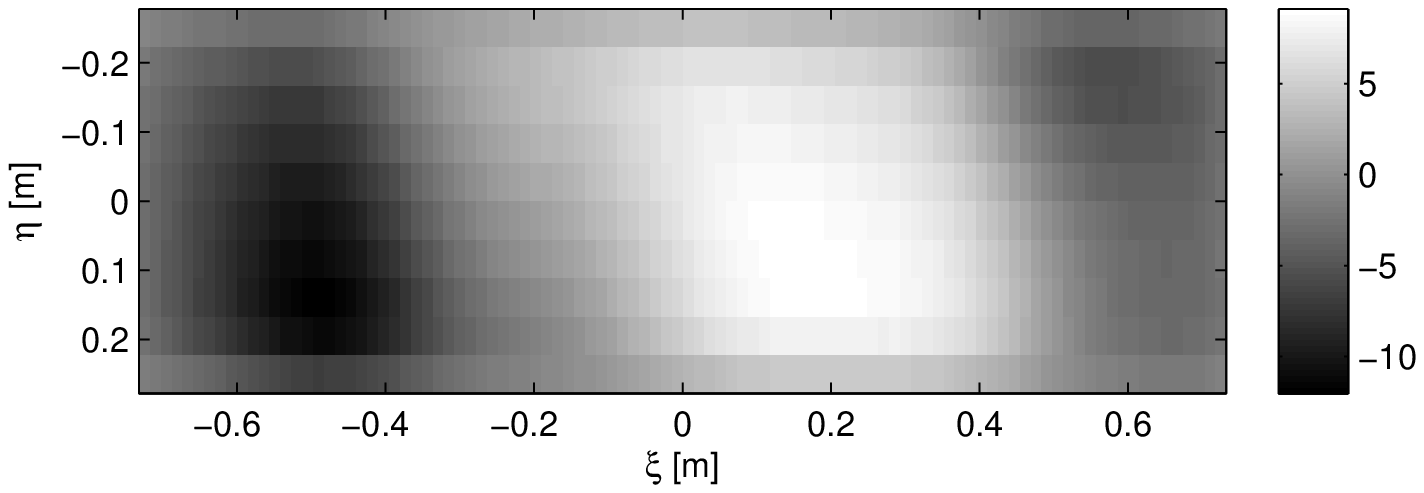}
\caption { Realistic WFE selected for case 6 (top) and the corresponding 
distribution of photo-centre (in $\mu$as) vs. patch position (bottom). } 
\label{fig_wfe_cog_c6}
\end{figure}

\subsubsection{Case 4} 
In case of symmetric aberration, i.e. changing the sign of the Zernike 
coefficients, the same RMS WFE is retained, and the focal plane image is 
symmetric with respect to case 1. 
Correspondingly, the photo-centre displacement, shown in figure 
\ref{fig_wfe_cog_c1_c4} (bottom), is opposite to the initial distribution 
of case 1: the mean and RMS are respectively -0.03 and 3.34~$\mu$as; 
the PTV is 14.13~$\mu$as. 
The photo-centre discrepancy between case 1 and case 4, after sign 
reversal of the latter, have both mean and RMS equal to zero within 
$10^{-6}\,\mu$as. 

\subsubsection{Case 5} 
The diffraction limit case is also worth considering, because the 
throughput non-uniformity introduces an asymmetry in the system, which 
is no longer ideal. 
This potentially might result in a photo-centre displacement. 
The simulation shows that this is not the case, since the photo-centre 
mean, RMS and PTV values are all zero. 
This is consistent with the mathematical framework of the diffraction integral, 
since replacing in Equation \ref{eq:ampli_mono} the null aberration $\Phi =0$, 
a unity pupil function is achieved, and we can see that sign reversal on the 
focal plane coordinates only introduces a global phase inversion on the complex 
argument, which vanishes in the square modulus providing an invariant intensity 
distribution (Equation \ref{eq:PSF_mono}). 
Besides, in mathematical terms, the Fourier transform $F(\omega)$ of a 
real function $f(t)$ is hermitian, so that $F(-\omega) = F^*(\omega)$ (complex 
conjugate), with the same result. 

\subsubsection{Case 6} 
\begin{figure}
\includegraphics[width=84mm]{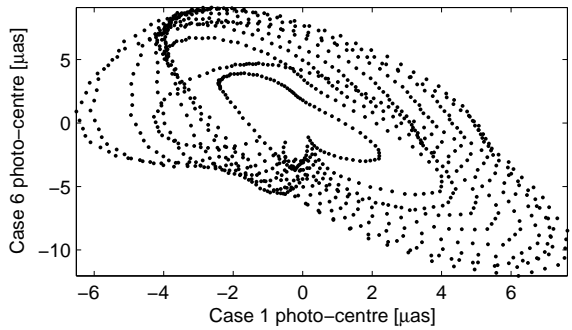}
\caption { Photo-centre distribution (in $\mu$as) of case 6 with respect 
to case 1. Different WFE distributions induce quite different astrometric 
errors. } 
\label{fig_cog_16}
\end{figure}

In order to test a different shape of aberration, an alternate set of Zernike 
coefficients is generated, in the same range of values, with a resulting 
RMS WFE of 35.23~nm. The same opacity patch distribution as above is used. 
The WFE distribution is shown in figure \ref{fig_wfe_cog_c6} (top); it is 
equivalent in terms of RMS value, but different in shape, to that of case 1 
(figure \ref{fig_wfe_cog_c1_c4}, top). 
The photo-centre displacement distribution, shown in figure \ref{fig_wfe_cog_c6} 
(bottom), has not only different shape with respect to that derived from the 
previous WFE distribution, but also larger deviations from the unperturbed 
case: the mean and RMS are respectively 0.13 and 5.45~$\mu$as; the PTV is 
21.14~$\mu$as. 
The plot of photo-centres of cases 1 and 6 against each other, shown in figure 
\ref{fig_cog_16}, in which each point is affected by the same opacity patch, 
does not evidence significant correlations. 
The RMS WFE is therefore not in itself a sufficient indication of the instrument 
sensitivity to non-uniform throughput variations. 

\subsubsection{Case 7} 
\begin{figure}
\includegraphics[width=84mm]{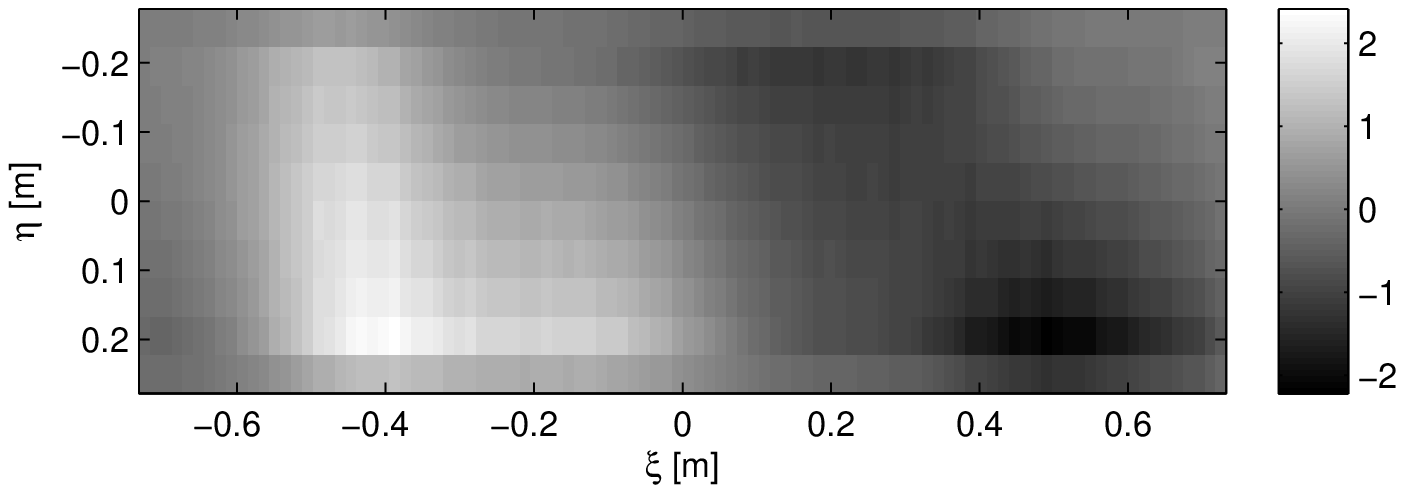}
\includegraphics[width=84mm]{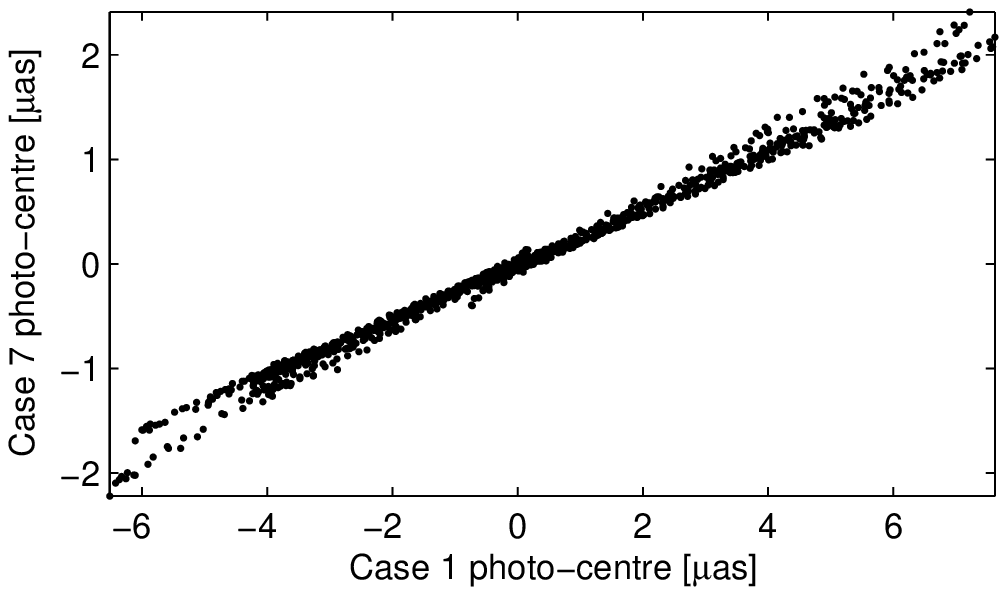}
\caption { Distribution of photo-centre (top, in $\mu$as) vs. patch position, 
case 7 (smaller patch), and photo-centre distribution (bottom, in $\mu$as) 
of case 7 with respect to case 1. } 
\label{fig_cog_c7}
\end{figure}

In this case, the characteristic size of the Gaussian distribution of 
opacity is reduced by a factor two, to $\sigma = 2.5\,$cm. 
Since the peak value remains set to 1\%, the effective perturbation to 
the system is smaller, and a reduced astrometric effect can be expected. 
The photo-centre displacement distribution, shown in figure \ref{fig_cog_c7} 
(top), is consistent with this expectation: the mean and RMS are respectively 
0.00 and 0.93~$\mu$as; the PTV is 4.63~$\mu$as. 
The photo-centre plot of case 7 vs. 1, shown in figure \ref{fig_cog_c7} 
(bottom), does evidence a significant correlation, since for 
each point the two cases have in common both the opacity patch and 
the WFE map. 
Scaling the former by a factor 3.56, defined by best fit, the 
discrepancy mean and RMS are respectively 0.02 and $0.29\,\mu$as. 
The scaling factor is close, but not equal, to the ratio of the opacity 
patch areas (i.e. four) on the pupil plane; given the non-linearity of 
the diffraction integral, Eqs. \ref{eq:ampli_mono} and \ref{eq:PSF_mono}, 
this seems to be reasonable. 

\subsection{Comparison} 
\begin{figure}
\includegraphics[width=84mm]{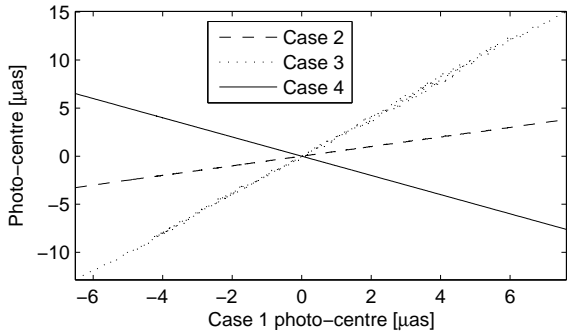}
\caption { Photo-centre distribution (in $\mu$as) of cases 2, 3 and 4 
with respect to case 1. Scaling the WFE amplitude induces a proportional 
change in the resulting astrometric error. } 
\label{fig_cog_1234}
\end{figure}

The various WFE cases are associated with different photo-centre 
displacements, induced by the same set of telescope throughput 
perturbations. 
A compact representation of the relationship among the first four cases is 
shown in figure \ref{fig_cog_1234}. 
The image profile is more aberrated in case 3 than in cases 1 and 2, and 
correspondingly a larger fraction of the total flux falls outside the 
readout region; this introduces additional noise. 
The astrometric error induced by a given transmission perturbation, 
if the WFE distribution is uniformly scaled by a factor 0.5, 2, and -1, 
is with good approximation multiplied by the same amount. 

Independent WFE distributions induce quite different astrometric errors, 
with respect to the same transmission perturbation, as shown in figure 
\ref{fig_cog_16}. 
Thus, the photo-centre displacement depends on the focal plane position. 

\section{Interferometry}
\label{sec_interf}
The opacity patch representing the non-uniform optical throughput of a 
telescope also has an impact on the performance of an interferometer. 
The orientation and phase distribution of the compressed beam is affected 
as derived in Sec. \ref{sec_1tel}, and this will contribute to the error 
budget of visibility and phase determination, depending on implementation 
aspects of the transport and combination optics \citep{Buscher08}. 
Additionally, a specific contribution to {\em baseline noise} can be 
identified. 

An effective pupil centre position can be defined for each opacity 
patch, as average weighted by the transmission distribution: 
\begin{equation}
\label{eq:pup_pos}
\xi_p  = 
{\int {d\xi d\eta \ \xi \cdot t (\xi, \eta)}
\over \int {d\xi d\eta \ t (\xi, \eta)}} 
=
\xi_0 {p_0 \over T_0} {A_p \over A_T - A_p} 
\, . 
\end{equation}
where $A_p$ and $A_T$ are the patch and primary mirror areas. 

Thus, the pupil centre displacement is related to the position of the 
opacity patch, weighted by the relative opacity vs. unperturbed transmission, 
and by the relative size of the patch vs. pupil. 

The opacity patch centre position $\xi_0$ can range over the whole geometric 
extension of the pupil, with uniform distribution, so that the average 
pupil displacement is zero: 
\begin{equation}
\label{eq:ave_disp}
\left\langle {\xi_p } \right\rangle  = \int {d\xi d\eta \ \xi _p }  
\equiv 0 \, , 
\end{equation}
and its RMS value is related to the RMS pupil size: 
\begin{equation}
\label{eq:rms_disp}
\sigma ^2 \left( {\xi _p } \right) = 
\left\langle {\left( {\xi _p  - \left\langle {\xi _p } \right\rangle } 
\right)^2 } \right\rangle  = 
\frac{{p_0 }} {{T_0 }} \frac{{A_p }} {{A_T  - A_p }} \xi_{RMS}^2 
\, . 
\end{equation}

In simple geometry cases, the RMS size of the aperture can be easily 
expressed in terms of the linear size $L_\xi$ in the relevant direction, 
for a rectangular pupil, or of the diameter $D$, for a circular unobstructed 
pupil, i.e. respectively 
$\xi_{RMS} = L_\xi / \sqrt{12}$ and $\xi_{RMS} = D/4$. 

Non-uniform variations of the optical throughput, therefore, induce 
apparent changes in the position of each telescope of an interferometric 
array. 

Let us consider the case of a two telescope interferometer with baseline $B$, 
observing a pair of stars with separation $\varphi << 1 \,$rad and zenithal 
distance $\theta$ (for the object with higher elevation), in a planar 
geometry for convenience. 
The variation of optical path difference (OPD) between the two stars, measured 
by the interferometer, is then $\Delta OPD \simeq \varphi \, B \, \cos \theta$, 
so that the baseline error $\sigma(B) = \sigma(\xi)$ is reflected into an 
astrometric error $\sigma(\varphi)$ on the determination of the star 
separation $\varphi$: 
\begin{equation}
\label{eq:err_phi}
\sigma \left( \varphi  \right) =  
\varphi {\sigma \left( B \right) \over B} 
\, . 
\end{equation}

Such effects are relevant e.g. for orbit determination, when measurements 
taken at different epochs $t_1,\,t_2$ are combined; in the mean time, the 
instrument may have suffered changes not easily identified (e.g. $<1\%$ 
in optical throughput) but sufficient to introduce relevant astrometric 
errors. 
We can evaluate the magnitude of the effect on two cases, namely 
PRIMA at the ESO VLTI (ref. PRIMA) and SIM. 

\subsubsection{PRIMA} 
The baseline is of order of 100~m, and the stellar separations considered 
for high precision astrometry in K band ($2.2\,\mu$m) are up to a few tens 
of arcsec. 
We consider a measurement performed by means of the Auxiliary Telescopes (AT, 
diameter 1.8~m), and assume one opacity patch with 1\% throughput degradation 
and size $\sigma = 5\,$cm, in an arbitrary position over the pupil. 
Then, the RMS baseline error is about 3.5~mm, and the corresponding error for 
baseline $B=100\,$m and separation $\varphi =10''$ is $\sigma(\varphi) =35\,\mu$as. 

\subsubsection{SIM}
The baseline is $B=6\,$m and the operating wavelength is in the visible, so 
that we select $\lambda = 600\,$nm; the individual apertures have diameter 
$D = 0.3\,$m. 
Narrow angle astrometry is performed on separations up to $\varphi  = 1^\circ$, 
so that, assuming an opacity patch with 1\% throughput degradation 
and size $\sigma = 1\,$cm, we get a RMS baseline error of 0.7~mm 
and angular measurement error of $\sigma(\varphi) =0''.4$. 

The larger astrometric error with respect to the PRIMA VLTI case is due to 
the combined contributions of shorter baseline and larger separation angle. 

\subsection{Calibration vs. metrology}
The above values of error introduced by non-uniform transmission variation 
are not immediately related to the final measurement performance, because 
diagnostics and correction are not yet considered. 
Of course, on-sky calibration procedures are foreseen, both for PRIMA and 
for SIM, in which the instrument is rapidly switched between the science 
target and suitable reference objects. 
The errors discussed above are systematic and strictly correlated (although 
not equal) among different sources, and this is a key factor in the 
definition of a measurement sequence leading to their suppression. 
Depending on availability of specific calibrators (e.g. binary systems or 
resolved objects considered sufficiently stable), simple calibration 
sequences can be devised. 

For example, if $\varphi_1$ and $\varphi_2$ are the angular quantities of 
interest referred to the science and reference sources, respectively, 
the measurement of the latter changes between epochs $t_1,\,t_2$ only 
because of instrument variation by $\sigma(\varphi_2)$, i.e. 
$\varphi_2 (t_2) = \varphi_2 (t_1) + \sigma(\varphi_2)$, 
whereas the former changes by both astrophysical and instrument amounts, 
respectively $\delta \varphi_1$ and $\sigma(\varphi_1)$: 
$\varphi_1 (t_2) = \varphi_1 (t_1) + \delta \varphi_1 + \sigma(\varphi_1)$. 
\\ 
From Equation \ref{eq:err_phi}, instrumental errors are correlated: 
\begin{equation}
\label{eq:err_phi_12}
\sigma \left( \varphi_1  \right) =  
\sigma \left( \varphi_2  \right) \, {\varphi_1 \over \varphi_2} 
\, , 
\end{equation}
so that the set of measurements allows correct determination of the 
desired astrophysical quantity $\delta \varphi_1$. 

In practice, the calibration procedure must be more complex, because 
the systematic errors affecting the measurement are likely to include 
additive components, as well as multiplicative components as 
from Equation \ref{eq:err_phi}. 

It should be noted that a metrology system installed to monitor the distance 
among fiducial points on the telescope structure is not sensitive to effective 
pupil displacements induced by non-uniform throughput variations, because 
the change involves optical parameters and not the instrument geometry. 
Metrology can be extremely useful with respect to short term disturbances 
from the environment, and to long term geometry perturbations, but 
is liable to failure in the detection of the effects discussed herein.

\section{Discussion} 
\label{sec_discuss}
The astrometric effects induced by non-uniform variations of the telescope 
throughput depend of course on the characteristic size and amplitude 
of the defects on the mirror surfaces. 
In turn, they depend on details of manufacturing and operating environment 
which are not easily identified nor measured. 
Useful practical indications may be provided by testing representative 
components e.g. during or after a significant period passed within an 
environment comparable to that of operation, as a cryogenic vacuum chamber, 
and subject to the expected radiation doses, as in \citet{Baccaro}.

\subsection{Design and implementation aspects}
For future astrometric instruments, using a comparably large field of view, 
the balancing of errors due to symmetric aberrations could be used to 
advantage, allowing compensation over a set of measurements taken in 
suitable conditions (i.e. cases 1 and 4 above). 
An aberration control approach could be conveniently adopted at the 
early design stage, minimising the sensitivity to manufacture tolerancing 
and alignment errors, also with particular regard to the optimisation 
criteria of in-orbit re-alignment, to preserve the symmetry. 
From a strictly experimental standpoint, it is possible to devise 
conceptual schemes for throughput measurement devices to be used on ground, 
for performance verification during the instrument integration, and even 
in orbit, to monitor the in-flight response. 
Their cost and benefit with respect to purely astronomical calibration 
is a matter of project trade-off. 

\subsection{Impact on Gaia}
For Gaia, a significant fraction of the astrometric effect 
associated with the image shape evolution induced by telescope throughput 
non-uniform variation {\em could} be taken care of within the framework 
of the self-calibration process, assuming that the instrument variation 
is sufficiently small over a time scale comparable with the coverage of 
the whole sky or a large part of it, e.g. order of six months. 
Also, part of the error may be averaged down among subsequent measurements, 
thus resulting in partial compensation. 

Partial compensation can be achieved also in case of symmetric aberrations 
over the field of view, as discussed in \citet{bus06} on the subject of 
chromatic astrometric errors. 
If opposite regions on the focal plane are affected by aberrations with 
opposite sign (as in cases 1 and 4 above), the astrometric errors introduced 
on the two measurements cancel out. 
Also in case of partial symmetry, or of limited balancing of the aberrations 
over the focal plane, some compensation can be achieved. 

Since the aberration distribution (and image profile) change over the focal 
plane, independently for each telescope, non-uniform throughput variations 
introduce astrometric errors in the large angle measurement. 
The distribution of photo-centre displacements in case of different WFE, 
as in figure \ref{fig_cog_16}, can provide an indication of the effect 
amplitude, but it is not fully representative because the patch positions 
are not correlated between the telescopes. 
The result is a field dependent error on the base angle, which is not 
strictly correlated to the measurements from the Base Angle Monitoring 
device (BAM), which probes only a small region of each optical system. 

An indication of throughput variation could be achieved by evaluation of the 
photometric response over a significant set of bright stars during each 
period of observation, in order to monitor the instrument response at the 
sub-milli-mag level (corresponding to 0.01\% overall transmission 
variation). 
However, photometry is not sufficient {\em per se} to provide adequate 
information, since it cannot distinguish the cases of uniform and 
non-uniform variation over the aperture. 

A wavefront sensor (WFS) of the Hartman-Shack type, splitting the collimated 
beam in several smaller regions, is potentially able to provide information 
on the variation of instrument transmission over its pupil, by photometric 
monitoring of the intensity from each region, as well as WFE derivative 
diagnostics by the elementary image displacements on the focal plane. 
However, a WFS is sensitive also to the variations of the non-common part 
of the system, namely the geometry and response of the lenslet array. 
As for photometry based diagnostics, the detector might also suffer 
changes which should be identified and factored out from the estimate of 
optical throughput variation. 

Besides, since the astrometric effect is associated with a variation of the 
effective signal profile, suitable image descriptors may be monitored, 
which usually have magnitude dependent sensitivity (e.g. image moments). 
The development of diagnostic tools based on image profile variation 
\citep{gai05} might prove useful to identification of critical instrument 
changes, in support to the previous approaches. 

Future investigations might tackle the development of efficient monitoring 
and correction algorithms, based on the available information, as well as 
the investigation of cumulative and field-dependent astrometric effects 
associated with the evolution of the telescope throughput.

\section{Conclusion} 
Non-uniform variations of the optical throughput over the telescope aperture, 
as experienced in ground based instruments, may provide astrometric effects 
significant at the level of several microarcseconds. 
This may in turn represent a non negligible contribution to the error 
budget of the most challenging astrometric experiments currently under 
implementation or proposed for the near future, as Gaia, PRIMA and SIM. 

The scale of the effects depends on manufacturing and operation parameters 
not easily measurable at the relevant level, i.e. order of 1\% on mirror 
reflectivity over regions of few cm, or 0.02\% on global throughput. 
Procedures of periodic calibration on sky can be used to retain 
acceptable levels of the systematic errors introduced by instrument 
response variations, in particular for long term measurements as 
interferometric orbits. 

For Gaia, the time scale of instrument evolution must be compared with 
the period of effective parameter calibration, to verify that the 
systematic errors are adequately removed from the data. 
Information from the WFS and from photometry might be exploited 
to monitor astrometric effects, as well as diagnostics of the image 
profile evolution.

\section*{Acknowledgments}
The analysis of interferometric effects was performed in the framework 
of the project PRIN INAF 2007 no. 6.


\label{lastpage}

\end{document}